\documentclass[a4paper, 8pt, italian, twocolumn]{article}
\usepackage{amsmath,amsfonts, amssymb, amsthm, epsfig}
\usepackage{babel}
\usepackage[latin1]{inputenc}
\begin{document}

\noindent{\bf SIMULTANEOUS NON-DISTURBING DETECTION
 OF INCOMPATIBLE PROPERTIES IN DOUBLE-SLIT EXPERIMENT}
\vskip.5pc\noindent
{$^{1,2}$ Giuseppe Nistic\`o and $^1$Angela Sestito}
\par\noindent
{\sl $^1$ Dipartimento di Matematica, Universit\`a della Calabria, Italy. \quad $^2$ Istituto Nazionale di Fisica Nucleare, Italy}\par\noindent
{\sl $^1$ Via Bucci 30b, 87036 Rende (CS) Italy, e-mail address: gnistico@unical.it}
\vskip.2in\noindent
We exploit the notion of which-slit detector introduced by
Englert, Scully and Walther (ESW), to show that two incompatible properties can be detected together for each particle hitting the screen,
without disturbing the center-of-mass motion of the particle.
\vskip.5pc
\noindent
{\bf 1.}\quad
Let us consider a typical two-slit experiment for a particle which possesses,
besides the position observable described in the Hilbert space $\cal K$,  a further
degree of freedom $s$ corresponding to a dichotomic
observable $S$ with spectrum $\sigma(S)=\{1,0\}$, such that $\partial H/\partial S={\bf 0}$ ($H$ is the Hamiltonian operator).
Such a system can be described
in the Hilbert
space ${\cal H}={\cal K}\otimes {\bf C}^2$.
The projection operator representing
the event `the particle crosses slit 1 (resp., 2)' has the form
$E=L\otimes{\bf 1}$ (resp., $({\bf 1}-L)\otimes{\bf 1}$). Given any interval $\Delta$ on the final screen,
by $F$ we denote the projection operator which represents the event
`the particle hits the final screen in a point within $\Delta$'.
Let $\psi_1\in\cal K$ and $\psi_0\in\cal K$ be state vectors respectively
localized in slit 1 and 2 when the particle is in the region of the two-slit, i.e. such that
\quad
$L\psi_1=\psi_1$, $L\psi_0=0$.
\par
If the complete state vector
of the particle is
$\Psi={(1/\sqrt 2)}[\psi_1\otimes\vert 1\rangle+\psi_0\otimes\vert 0\rangle]$, then
the projection $T={\bf 1}\otimes\vert 1\rangle\langle 1\vert$
represents a which-slit detector which does not disturb the center-of-mass
motion of the particle, because\par\noindent
\quad $[T,F]={\bf 0}$\hfill{(1)}\par\noindent
\quad $[T,E]={\bf 0}$ and $T\Psi=E\Psi$.\hfill{(2)}\par
\par\noindent
Indeed, (1) implies that the measurement of $T$ can be performed without affecting the distribution
of the particles on the final screen; (2) means that $T$ and $E$ represent directly correlated properties,
so that the outcome 1 (resp., 0) for $T$ detects the passage of the particle through slit 1 (resp.2).
Since $T$ is {\sl completely} non-disturbing, it is an idealized version
of the which-slit detector introduced by ESW [1][2].\par
The following definition generalizes this notion.
\par\noindent
DEFINITION.
{\sl Let $L_+$ be a projection of ${\cal K}$.
A projection $T_+={\bf 1}\otimes R$ is called a non disturbing ESW detector for $E_+=L_+\otimes{\bf 1}$ if
\par\noindent
\quad $[T_+,F]={\bf 0}$\hfill{(3)}\par\noindent
\quad $[T_+,E_+]={\bf 0}$ and $T_+\Psi=E_+\Psi$.\hfill{(4)}}
\par\noindent
Hence, projection $T$ above is a non-disturbing (which-slit) ESW
detector for projection $E$.\par
 \vbox{\vskip.35in}\noindent
{\bf 2.}\quad Now we face the following question: {\sl There are
situations in which the simultaneous knowledge of which slit
passage and an incompatible property is possible for each particle
hitting the final screen, without affecting the distribution of
particles?} As we show by means of the following example, the
answer is {\sl yes}, and in such a case the two properties are
directly correlated.
\par
Let $M_1$, $M_2$, $M_3$, $M_4$ be four mutually
orthogonal subspaces of $\cal K$, with respective projections $P_1$,
$P_2$, $P_3$, $P_4$, such that
\quad
$L=P_1+P_2$,\quad ${\bf 1}-L=P_3+P_4$.
\par\noindent
Let the state of the particle be described by
$$
\Psi={(1/2)}\{(\psi_1+\psi_2)\vert 1\rangle+(\psi_3+\psi_4)\vert 0\rangle\},
$$
where $\psi_k\in M_k$ and $\Vert\psi_k\Vert=1$,
$k=1,2,3,4$,.
Since (1) and (2) continue to hold,
projection $T$ above is a non-disturbing ESW which-slit detector.
\par\noindent
Now we consider the projection $E_+=L_+\otimes{\bf 1}$,
where
\par\noindent
$L_+=A+B+C+D$, with
\par\noindent
$A=\vert\psi_1\rangle[(3/4)\langle\psi_1\vert+{(1/4)}\langle \psi_2\vert
-{(1/4)}\langle\psi_3\vert+{(1/4)}\langle \psi_4\vert]$,
\par\noindent
$B=\vert\psi_2\rangle[{(1/4)}\langle\psi_1\vert+{(3/4)}\langle \psi_2\vert
+{(1/4)}\langle\psi_3\vert -{(1/4)}\langle \psi_4\vert]$,
\par\noindent
$C=\vert\psi_3\rangle[-{(1/4)}\langle\psi_1\vert+{(1/4)}\langle \psi_2\vert
+{(1/4)}\langle\psi_3\vert -{(1/4)}\langle \psi_4\vert]$,
\par\noindent
$D=\vert\psi_4\rangle[{(1/4)}\langle\psi_1\vert-{(1/4)}\langle \psi_2\vert
-{(1/4)}\langle\psi_3\vert+{(1/4)}\langle \psi_4\vert]$.
\par\noindent
To grasp the physical meaning of $E_+$ the choice of $\psi_1,\psi_2,\psi_3,\psi_4$ must be specified,
of course. However, rule
$[E_+,E]\neq{\bf 0}$ holds for any choice; therefore {\sl projection
$E_+$ represents a  property incompatible with $E$}.
Now, equality $T\Psi=E_+\Psi$ holds; hence $T$ is a non-disturbing ESW detector also for $E_+$, which is incompatible with
$E$.
\par
Furthermore, since $E\Psi=T\Psi=E_+\Psi$, when the particles are
prepared in the state $\Psi$, there is an entanglement between the
two incompatible properties $E$ and $E_+$. \vskip.5in
 \epsfbox[-30 30 150 80]{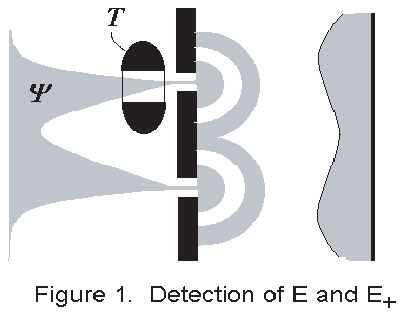}
 \vskip.5in
\noindent Outcome $1$ ($0$) for which slit detector $T$ implies
that both $E$ and $E_+$ have outcome $1$ ($0$), $E$ and $E_+$
being incompatible with each other. The measurement of $T$ does
not affect the final position. \vskip.3in
 \noindent [1] M.O. Scully, B-G.
Englert, H. Walther, (1991), {\sl Nature}, {\bf 351}, 111-116
\par\noindent
[2] G. Nistic\`o, M.C. Romania, (1994), {\sl J.Math.Phys.}, {\bf
35}, 4534-4546
\end{document}